\documentclass[twocolumn,aps,prl,showpacs,superscriptaddress]{revtex4-1}
\usepackage[dvipdf]{graphicx}
\usepackage{slashed}
\usepackage{bm}
\begin{document}
\title{A relativistic electron in an anisotropic conduction band}
\author{Aiying Zhao}
\email{ayzhao0909@knights.ucf.edu}
\affiliation{Department of Physics, University of Science and Technology Beijing, 30 Xueyang Road, Haidian District, Beijing 100083, China}
 \affiliation{Department of Physics, University of Central Florida, 4111 Libra Drive, Orlando, FL 32816-2385 USA}
 \author{Jingchuan Zhang}
 \email{zhangjingchuan@outlook.com}
\affiliation{Department of Physics, University of Science and Technology Beijing, 30 Xueyang Road, Haidian District, Beijing 100083, China}
 \affiliation{Department of Physics, University of Central Florida, 4111 Libra Drive, Orlando, FL 32816-2385 USA}
\author{Qiang Gu}\email{qgu@ustb.edu.cn, corresponding author}\affiliation{Department of Physics, University of Science and Technology Beijing, 30 Xueyang Road, Haidian District, Beijing 100083, China}\author{Richard A. Klemm}
\email{richard.klemm@ucf.edu, corresponding author} \affiliation{Department of Physics, University of Central Florida, 4111 Libra Drive, Orlando, FL 32816-2385 USA}
\date{\today}
\begin{abstract}
The  Dirac equation is extended for a relativistic electron  in an orthorhombically-anisotropic conduction band.  Its covariance is established with general proper and improper Lorentz transformations.  In the  non-relativistic limit,  the kinetic  and  Zeeman energy terms of the Hamiltonian are both determined by the same three effective masses, and the quantum spin Hall effect is derived.  This has important consequences for magnetic measurements   of many classes of clean anisotropic semiconductors, metals, and superconductors.  The Zeeman energy is vanishingly small for magnetic fields parallel to clean monolayers and  in all directions in  quasi-one-dimensional materials. \end{abstract}
 \vskip0pt\vskip0pt
\maketitle
\section{Introduction}
In semiconductors such as Si and Ge, and in metallic Bi, the lowest energy conduction bands have ellipsoidal symmetry, with $\epsilon_0({\bm p})=\sum_{i=1}^3p_i^2/(2m_i)$ about some point in the first Brillouin zone.  \cite{Cohen,CohenBlount,Mahan}  The normal state of the high-temperature superconductor YBa$_2$CuO$_{7-\delta}$ is metallic in all three orthorhombic directions \cite{Friedmann,Klemmbook}, so that it is reasonable to assume its $\epsilon_0({\bm p})$ has that form, with $m_3\gg m_1, m_2$.  But as the planar $^{63}$Cu Knight shift measurements on that material showed anomalous temperature-independent behavior through the superconducting state for the applied magnetic field ${\bm H}||\hat{\bm c}$ \cite{Barrett}, and the apparent incompatibility of Knight shift measurements with scanning tunneling measurements of the superconducting gap in Sr$_2$RuO$_4$ \cite{Ishida,Suderow}, the need for an improved theory of the Knight shift that incorporates the orbital and spin components of the motion of the conduction electrons is sorely needed \cite{HallKlemm,magnetochemistry}.  In addition, upper critical field $H_{c2,||}(T)$ measurements of superconducting single-layer doped transition metal dichalcogenides  and possibly also of twisted graphene bilayers provided evidence for an apparent violation of the Pauli limit for the applied field parallel to the superconducting layer or bilayers \cite{NbSe2,MoS2a,MoS2b,WTe2a,WTe2b,graphene2}, even though the standard theory assumed such violation to strong spin-orbit scattering \cite{KLB}.

 Since  such materials can now be prepared in the clean limit \cite{graphene2}, it is important to reassess the  effect of the crystal anisotropy upon the Landau orbits and Zeeman energies.  Here we present a theory of a relativistic electron in an orthorhombically anisotropic conduction band, and demonstrate its covariance by its invariance  under the most general proper Lorentz transformation,  and argue for its invariance under the improper transformations of parity, charge conjugation, and time reversal.  The Foldy-Wouythuysen transformation is used  to evaluate the non-relativistic limit of this anisotropic Hamiltonian to order $(mc^2)^{-2}$. In two spatial dimensions, the Zeeman interaction only exists for the magnetic induction ${\bm B}$ normal to the conducting plane.    In one dimension, there is no Zeeman interaction.  These results lead to the absence of Pauli limiting of $H_{c2,||}(T)$ in clean ultrathin superconductors, as  observed recently \cite{NbSe2,MoS2a,MoS2b,WTe2a,WTe2b,graphene2}.  They also have profound implications for  interpretations of Knight shift measurements on highly anisotropic superconductors such as Sr$_2$RuO$_4$ and (TMTSF)$_2$PF$_6$ \cite{Ishida,Chaikin,HallKlemm,magnetochemistry}.

 For a relativistic electron of rest mass $m$ and charge $-|e|$, the kinetic energy part of the Hamiltonian is given by $T=\sqrt{({\bm p}+e{\bm A})^2c^2+m^2c^4}$, where  ${\bm A}$ is the magnetic vector potential. Dirac first noted that with ${\bm p}\rightarrow -i\hbar{\bm\nabla}$, the kinetic energy part of the Schr{\"o}dinger equation $T\psi=i\hbar\frac{\partial\psi}{\partial t}$,  when squared, $T^2\psi=-\hbar^2\frac{\partial^2\psi}{\partial t^2}$, recovered the  Klein-Gordon equation, the wave equation for a free particle of charge $-|e|$ and mass $m$ \cite{Dirac,BjorkenDrell}.   He then showed that it could also be recovered by choosing the linearized Hamiltonian ${\cal H}$ of the form \cite{Dirac,BjorkenDrell}
\begin{eqnarray}
i\hbar\frac{\partial\psi}{\partial t}= \Bigl[c{\bm\alpha}\cdot\Bigl(\frac{\hbar}{i}{\bm\nabla}+e{\bm A}\Bigr)+\beta mc^2\Bigr]\psi\equiv {\cal H}\psi.
\end{eqnarray}
 Dirac required $\left\{\alpha^{\mu},\alpha^{\nu}\right\}=2\delta^{\mu\nu}$ in contravariant form,  $\left\{\alpha^{\mu},\beta\right\}=0$, $\beta^2=1$, and  for three spatial dimensions,   the $\alpha^{\mu}$ and $\beta$ to be the traceless rank 4 matrices
\begin{eqnarray}
\alpha^{\mu}&=&\left[\begin{array}{cc}0&\sigma^{\mu}\\
\sigma^{\mu}&0\end{array}\right],\hskip20pt \beta=\left[\begin{array}{cc}1&0\\
0&-1\end{array}\right],\label{beta}
\end{eqnarray}
 where the $\sigma^{\mu}$ are the Pauli matrices, and $1$ implies the rank 2 identity matrix\cite{Dirac,BjorkenDrell}.
\section{The model}
  However, for the  conduction electrons in a large number of semiconductors, metals, and superconductors, the single particle states in the primary conduction band or bands are anisotropic, and the simplest expression for the anisotropic relativistic kinetic energy $\tilde{T}$ may be written for orthorhombic anisotropy as
\begin{eqnarray}
\tilde{T}&=&\sqrt{\sum_{i=1}^3m (p_i+eA_i)^2c^2/m_i+m^2c^4},
\end{eqnarray}
where  $m_i$  is the effective mass of the electron in the $i^{\rm th}$ direction,  $\hat{\bm e}_i=\hat{\bm x},\hat{\bm y},\hat {\bm z}$, and $m$ is the electron rest mass.  For electrons in a conduction band of a semiconductor, metal, or superconductor, $mc^2$ is the large energy in $\tilde{\cal H}$, so that $\sum_{i=1}^3(p_i+eA_i)^2/m_i\ll mc^2$.
The question of how to treat the combined effects of the interactions of an electron with an applied electric ${\\bm E}$ and magnetic ${\bm B}={\bm\nabla}\times{\bm A}$ fields while in an anisotropic medium has long been of interest.  Here we derive a fully relativistic approach to the problem of an electron in an anisotropic conduction band that provides the correct, self-consistent form for the interaction of the conduction electrons with a external fields in an anisotropic semiconductor, metal, or superconductor.

In this anisotropic case, a relativistic electron in an anisotropic environment of orthorhombic symmetry satisfies the Schr{\"o}dinger equation based upon the modified Hamiltonian $\tilde{\cal H}=\tilde{T}+V$, where the electrostatic potential energy  $V({\bm r})=-e\Phi({\bm r})$ due to the electrostatic potential $\Phi({\bm r})$ is included, \cite{BjorkenDrell}
\begin{eqnarray}
i\hbar\frac{\partial\psi}{\partial t}&=&\Bigl[c\tilde{\bm \alpha}\cdot\Bigl(\frac{\hbar}{i}{\bm\nabla}+e{\bm A}\Bigr)+\beta mc^2-e\Phi\Bigr]\psi=\tilde{\cal H}\psi,\label{contravariantDirac}
\end{eqnarray}
where $\beta$ is given in Eq. (\ref{beta}),
\begin{eqnarray}
\tilde{\alpha}^{\mu}&=&\left[\begin{array}{cc}0&\tilde{\sigma}^{\mu}\\
\tilde{\sigma}^{\mu}&0\end{array}\right],\hskip20pt\tilde{\sigma}^{\mu}=\frac{\sigma^{\mu}}{\sqrt{\tilde{m}_{\mu}}}
\end{eqnarray} for $\mu=1,2,3$, where
$\tilde{m}_{\mu}=m_{\mu}/m$.  We used subscripts for the contravariant forms of the effective masses, in order to avoid confusion with superscripts representing exponents, which appear subsequently.  The matrices satisfy $\left\{\tilde{\alpha}^{\mu},\tilde{\alpha}^{\nu}\right\}=2\delta^{\mu\nu}/\tilde{m}_{\mu}$, and $\left\{\tilde{\alpha}^{\mu},\beta\right\}=0$.

  From Eq. (\ref{contravariantDirac}), it is easy to show the $\mu^{\rm th}$ component of the probability  current for $\mu=1, 2, 3$ is   $j^{\mu}=\psi^{\dagger}\tilde{\alpha}^{\mu}\psi$, and since $\rho=\psi^{\dag}\psi$, the continuity equation
$\frac{\partial\rho}{\partial t}+\frac{\partial}{\partial x^{\mu}}j^{\mu}=\frac{\partial\rho}{\partial t}+{\rm div}{\bm j}=0$,
 is still satisfied with effective mass anisotropy.

\section{Covariant anisotropic Dirac equation}
 In order to demonstrate the Lorentz invariance of this anisotropic Dirac equation, we multiply its contravariant form, Eq. (\ref{contravariantDirac}), by $\beta$,
\begin{eqnarray}
\tilde{\gamma}^0&=&\beta=\left[\begin{array}{cc}1&0\\0&-1\end{array}\right],\hskip15pt\tilde{\gamma}^{\mu}=\beta\tilde{\alpha}^{\mu}=\left[\begin{array}{cc}0&\tilde{\sigma}^{\mu}\\ -\tilde{\sigma}^{\mu}&0\end{array}\right],\>\>\>\>
\end{eqnarray}
for $\mu=1,2,3$.  The $\tilde{\gamma}^{\mu}$ for $\mu=1,2,3$ satisfy the anticommutation relations
\begin{eqnarray}
\left\{\tilde{\gamma}^{\mu},\tilde{\gamma}^{\nu}\right\}&\equiv&2g^{\mu\nu}\delta^{\mu\nu}=\frac{-2\delta^{\mu\nu}}{\tilde{m}_{\mu}}
\end{eqnarray}
where $\tilde{\gamma}^0$ is hermitian so that $(\tilde{\gamma}^0)^2=1$, the $\tilde{\gamma}^{\mu}$ for $\mu=1,2,3$ are antihermitian, and the metric $\tilde{g}$ is given by
\begin{eqnarray}
\tilde{g}&=&\left(\begin{array}{cccc}1&0&0&0\\
0&-\tilde{m}_1^{-1}&0&0\\
0&0&-\tilde{m}_2^{-1}&0\\
0&0&0&-\tilde{m}_3^{-1}\end{array}\right).\label{tildeg}
\end{eqnarray}
We then may use the Feynman slash notation \cite{BjorkenDrell},
\begin{eqnarray}
\tilde{\slashed{\nabla}}&=&\tilde{\gamma}^{\mu}\frac{\partial}{\partial x^{\mu}}=\frac{\tilde{\gamma}^0}{c}\frac{\partial}{\partial t}+\tilde{\bm\gamma}\cdot{\bm\nabla},\nonumber\\
\tilde{\slashed{A}}&=&\tilde{\gamma}^{\mu}A_{\mu}=\tilde{\gamma}^0A^0-\tilde{\bm\gamma}\cdot{\bm A},
\end{eqnarray}
to write the anisotropic Dirac equation in covariant form,
\begin{eqnarray}
(i\hbar\tilde{\slashed{\nabla}}+e\tilde{\slashed{A}}-mc)\psi&=&0.\label{covariantDirac}
\end{eqnarray}

 In the following, we demonstrate that the norm for a relativistic electron in an orthorhombically anisotropic conduction band with metric $\tilde{g}$ given by Eq. (\ref{tildeg}) is indeed invariant under the most general proper Lorentz transformation ${\tilde A}$, find the matrix form of ${\tilde{A}}$, and show that it  has O(3,1) symmetry, as for the isotropic case. Examples of  general rotations and general boosts are given. Improper Lorentz transformations such as reflections, parity, charge conjugation,  and time inversion can be treated exactly as for the isotropic Dirac equation \cite{BjorkenDrell,Jackson}.
\section{Proof of covariance}
For a general proper Lorentz transformation in a relativistic orthorhombic system,
$x'=\tilde{a}x$, where $x'$ and $x$ are column  (Nambu) four-vectors and $\tilde{a}$ is the appropriate proper anisotropic Lorentz transformation, which is to be found based upon symmetry arguments.  We require the norm with $\tilde{g}$ to be invariant under all possible Lorentz transformations \cite{Jackson}
$(x,\tilde{g}x)=(x',\tilde{g}x')$,
or
$x^{T}\tilde{g}x=(x')^{T}\tilde{g}x'$,
where $x^T$ is the transpose (row) form of the four-vector $x$ and  $\tilde{g}$ is given by Eq. (8).
We then have
$x^T\tilde{g}x=(x')^T\tilde{g}x'=x^T\tilde{a}^T\tilde{g}\tilde{a}x$,
 which implies
$\tilde{g}=\tilde{a}^T\tilde{g}\tilde{a}$.
As for the isotropic case, we assume
$\tilde{a}=e^{\tilde{L}}$,
so that
$\tilde{a}^T=e^{\tilde{L}^T}$, and
$\tilde{a}^{-1}=e^{-\tilde{L}}$.
Then from $\tilde{g}=\tilde{a}^T\tilde{g}\tilde{a}$, we have $\tilde{g}\tilde{a}^{-1}=\tilde{a}^T\tilde{g}$ and hence that $\tilde{a}^{-1}=\tilde{g}^{-1}\tilde{a}^T\tilde{g}$.  We then may rewrite this as
\begin{eqnarray}
e^{-\tilde{L}}&=&\tilde{g}^{-1}e^{\tilde{L}^T}\tilde{g}=e^{\tilde{g}^{-1}\tilde{L}^T\tilde{g}}.
\end{eqnarray}
 Taking the logarithm of both sides, we obtain
$-\tilde{L}=\tilde{g}^{-1}\tilde{L}^T\tilde{g}$, or that
$-\tilde{g}\tilde{L}=\tilde{L}^T\tilde{g}=(\tilde{g}\tilde{L})^T$,
which requires  $\tilde{g}\tilde{L}$ to be  antisymmetric.  We then write \cite{Jackson}
\begin{eqnarray}
\tilde{L}&=&\left(\begin{array}{cccc}0&\frac{-\zeta_1}{\sqrt{\tilde{m}_1}}&\frac{-\zeta_2}{\sqrt{\tilde{m}_2}}&\frac{-\zeta_3}{\sqrt{\tilde{m}_3}}\\
&&&\\
-\zeta_1\sqrt{\tilde{m}_1}&0&\frac{\omega_3\sqrt{m_1}}{\sqrt{m_2}}&\frac{-\omega_2\sqrt{m_1}}{\sqrt{m_3}}\\
&&&\\
-\zeta_2\sqrt{\tilde{m}_2}&\frac{-\omega_3\sqrt{m_2}}{\sqrt{m_1}}&0&\frac{\omega_1\sqrt{m_2}}{\sqrt{m_3}}\\
&&&\\
-\zeta_3\sqrt{\tilde{m}_3}&\frac{\omega_2\sqrt{m_3}}{\sqrt{m_1}}&\frac{-\omega_1\sqrt{m_3}}{\sqrt{m_2}}&0\end{array}\right),\nonumber\\
\end{eqnarray}
for which $\tilde{g}\tilde{L}$ is easily shown to be antisymmetric.

We may then write
\begin{eqnarray}
\tilde{L}&=&-{\bm\omega}\cdot{\tilde{\bm S}}-{\bm\zeta}\cdot{\tilde{\bm K}},
\end{eqnarray}
where
\begin{eqnarray}
\tilde{K}_1&=&\left(\begin{array}{cccc}0&\tilde{m}_1^{-1/2}&0&0\\
\tilde{m}_1^{1/2}&0&0&0\\
0&0&0&0\\
0&0&0&0\end{array}\right),\\
\tilde{K}_2&=&\left(\begin{array}{cccc}0&0&\tilde{m}_2^{-1/2}&0\\
0&0&0&0\\
\tilde{m}_2^{1/2}&0&0&0\\
0&0&0&0\end{array}\right),\\
\tilde{K}_3&=&\left(\begin{array}{cccc}0&0&0&\tilde{m}_3^{-1/2}\\
0&0&0&0\\
0&0&0&0\\
\tilde{m}_3^{1/2}
&0&0&0\end{array}\right),\\
\tilde{S}_1&=&\left(\begin{array}{cccc}0&0&0&0\\
0&0&0&0\\
0&0&0&-\sqrt{\frac{m_2}{m_3}}\\
0&0&\sqrt{\frac{m_3}{m_2}}&0\end{array}\right),\\
\tilde{S}_2&=&\left(\begin{array}{cccc}0&0&0&0\\
0&0&0&\sqrt{\frac{m_1}{m_3}}\\
0&0&0&0\\
0&-\sqrt{\frac{m_3}{m_1}}&0&0\end{array}\right),\\
\tilde{S}_3&=&\left(\begin{array}{cccc}0&0&0&0\\
0&0&-\sqrt{\frac{m_1}{m_2}}&0\\
0&\sqrt{\frac{m_2}{m_1}}&0&0\\
0&0&0&0\end{array}\right).
\end{eqnarray}
It is easy to show that
$\left[\tilde{S}_i,\tilde{S}_j\right]=\epsilon_{ijk}\tilde{S}_k$,
$\left[\tilde{K}_i,\tilde{K}_j\right]=-\epsilon_{ijk}\tilde{S}_k$, and
$\left[\tilde{S}_i,\tilde{K}_j\right]=\epsilon_{ijk}\tilde{K}_k$,
so the anisotropic Lorentz transformation matrix $\tilde{L}$ has SL(2,C) or O(3,1) group symmetry, precisely as for the isotropic case \cite{Jackson}.
We now  show some  examples.  We first define
$\omega=\sqrt{\omega_1^2+\omega_2^2+\omega_3^2}$ and then write
\begin{eqnarray}
A_i&=&\cos\omega+\frac{\omega_i^2}{\omega^2}(1-\cos\omega),\\
B_{ijk}^{\pm}&=&\Bigl(\frac{m_i}{m_j}\Bigr)^{1/2}\Bigl[\frac{\omega_i\omega_j}{\omega^2}(1-\cos\omega)\pm\frac{\omega_k}{\omega}\sin\omega\Bigr].
\end{eqnarray}
Then, for the general rotation case, we have
\begin{eqnarray}
e^{-{\bm\omega}\cdot{\tilde{\bm S}}}&=&
\left(\begin{array}{cccc}1&0&0&0\\
0&A_1&B^{+}_{123}&B^{-}_{132}\\
0&B_{213}^{-}&A_2&B^{+}_{231}\\
0&B^{+}_{312}&B^{-}_{321}&A_3\end{array}\right),
\end{eqnarray}
the determinant of which is 1, as required for a rotation.

For the general boost case, we first set ${\bm \zeta}=\hat{\overline{\bm \beta}}\tanh^{-1}\overline{\beta}$, where $\overline{\bm\beta}={\bm v}/c$, ${\bm v}$ is the electron's velocity, and define
$\zeta=\sqrt{\zeta_1^2+\zeta_2^2+\zeta_3^2}$,
$\cosh\zeta=\gamma=\frac{1}{\sqrt{1-\overline{\beta}^2}}$,
$\sinh\zeta=\gamma\overline{\beta}$, and
$\overline{\beta}=\sqrt{\overline{\beta}_1^2+\overline{\beta}_2^2+\overline{\beta}_3^2}$, as for the isotropic case \cite{Jackson}.  Then we define
\begin{eqnarray}
C_{i}^{\pm}&=&-\gamma\overline{\beta}_i\tilde{m}_i^{\pm 1/2},\\
D_{i}&=&1+\frac{(\gamma-1)\overline{\beta}_i^2}{\overline{\beta}^2},\\
E_{ij}&=&(\gamma-1)\frac{\overline{\beta}_i\overline{\beta}_j}{\overline{\beta}^2}\Bigl(\frac{m_i}{m_j}\Bigr)^{1/2}.
\end{eqnarray}
Then for the general boost case, we have \cite{Jackson}
\begin{eqnarray}
e^{-{\bm\zeta}\cdot\tilde{\bm K}}
&=&\left(\begin{array}{cccc}\gamma&C_1^{-}&C_2^{-}&C_3^{-}\\
C_1^{+}&D_1&E_{12}&E_{13}\\
C_2^{+}&E_{21}&D_2&E_{23}\\
C_3^{+}&E_{31}&E_{32}&D_3\end{array}\right),
\end{eqnarray}
the determinant of which is  also 1,
as required.

Hence the local anisotropic Dirac equation in its covariant form, Eq. (\ref{covariantDirac}), is invariant under the most general proper Lorentz transformation.  As argued in the following, it is also invariant under all of the relevant improper Lorentz transformations:  reflections or parity, charge conjugation, and time reversal.  Those operations are precisely the same as for the isotropic covariant Dirac equation \cite{BjorkenDrell}.

Improper Lorentz transformations are represented by rank-4 matrices $b$ satisfying ${\rm det}(b)=-1$.  As for the isotropic case, reflections require ${\bm x}'=-{\bm x}$ and $t'=t$, so that $b$ is a diagonal rank-4 matrix with $b^{00}=1$ and $b^{\mu\mu}=-1$ for $\mu=1,2,3$, which is identical to $\tilde{g}$ in the isotropic limit $\tilde{m}_{\mu}\rightarrow1$ for $\mu=1,2,3$ \cite{BjorkenDrell}. Reflections can then be represented by a unitary  matrix $P$ satisfying
\begin{eqnarray}
P^{-1}\tilde{\gamma}^{\mu}P&=&b^{\mu\mu}\tilde{\gamma}^{\mu},
\end{eqnarray}
which is satisfied for
\begin{eqnarray}
P&=&e^{i\phi}\tilde{\gamma}^0,
\end{eqnarray}
where the phase factor $\phi=n\pi/2$ for integer $n$, so that four reflections leaves $\psi$ invariant, as for a rotation through $4\pi$ about the quantization axis of a spin 1/2 spinor.  We also have that $P\Phi({\bm x},t)=\Phi'({\bm x}',t)=\Phi({\bm x},t)$ is even under parity, and $P{\bm A}({\bm x},t)={\bm A}'({\bm x}'.t)=-{\bm A}({\bm x},t)$ is odd under parity.  This is exactly the same as for the isotropic Dirac equation.

With regard to charge conjugation, the hole wave function $\psi_c$ in an anisotropic conduction band  satisfies
\begin{eqnarray}
(i\hbar\tilde{\slashed{\nabla}}-e\tilde{\slashed{A}}-mc)\psi_c&=&0.\label{positronDirac}
\end{eqnarray}
As for the isotropic Dirac equation, this is accomplished by taking the complex conjugate:
\begin{eqnarray}
\psi_c&=&C\overline{\psi}^T,
\end{eqnarray}
where $C=i\gamma^2\gamma^0$. As for the isotropic Dirac equation, charge conjugation in the  anisotropic Dirac equation is also given by  $C=i\gamma^2\gamma^0$, so that
\begin{eqnarray}
C^{-1}\tilde{\gamma}^{\mu}C&=&-\tilde{\gamma}^{\mu T}.
\end{eqnarray}
This works precisely the same as for the isotropic Dirac equation.

As for time reversal, we have the properties that if $t'=-t$, we have
\begin{eqnarray}
\psi'(t')&=&T\psi^{*}(t),
\end{eqnarray}
where for the isotropic Dirac equation,
\begin{eqnarray}
T&=&i\gamma^1\gamma^3.
\end{eqnarray}
For the anisotropic Dirac equation, the same definition of $T$ applies, so the invariance under time reversal is equivalently established.
\section{Non-relativistic Hamiltonian}
We used   the Foldy-Wouthuysen transformations for the anisotropic operator ${\cal O}=\tilde{\bm\alpha}\cdot({\bm p}+e{\bm A})$ to obtain the non-relativistic limit of Eq. (\ref{covariantDirac}) \cite{BjorkenDrell}. We find to order $(mc^2)^{-2}$ that relative to $\beta mc^2$, the Hamiltonian becomes
\begin{eqnarray}
{\cal H}^{NR}&=&\sum_{i=1}^3\Bigl[\beta\Bigl(\frac{\Pi_i^2}{2m_i}
+\frac{\mu_B\sqrt{\tilde{m}_i}\sigma_iB_i}{(\tilde{m}_g)^{3/2}}\Bigr)+\frac{\mu_B}{4mc^2}\frac{1}{\tilde{m}_i}\frac{\partial E_i}{\partial x_i}\Bigr)\Bigr],\nonumber\\
& &+\frac{\mu_B}{4mc^2(\tilde{m}_g)^{3/2}}\sum_{\mu,\nu,\lambda=1}^3\Bigl[\Bigl(2E_{\mu}(p_{\nu}+eA_{\nu})+i\frac{\partial E_{\mu}}{\partial x_{\nu}}\Bigr)\nonumber\\
& &\hskip90pt\times\epsilon_{\mu\nu\lambda}\sqrt{\tilde{m}_{\lambda}}\sigma_{\lambda}\Bigr]-e\Phi,\label{HNR}
\end{eqnarray}
where $\Pi_i=p_i+eA_i$, the geometric mean $\tilde{m}_g=(\tilde{m}_1\tilde{m}_2\tilde{m}_3)^{1/3}$, $\mu_B=\frac{e\hbar}{2m}$ is the Bohr magneton in SI units, $\epsilon_{\mu\nu\lambda}$ is the Levi-Civita symbol, and $m_i=m\tilde{m}_i$. In this limit, the Hamiltonian for an electron is obtained with $\beta=1$.  Although not included in that derivation \cite{BjorkenDrell}, ${\bm A}$ arises in this order, and leads to the quantum spin Hall effect.

In the isotropic 3D limit  with $\tilde{m}_i=1$,
  Eq. (\ref{HNR}) reduces   to
\begin{eqnarray}
{\cal H}_{3D}^{NR}&=&\frac{{\bm\Pi}^2}{2m}+\mu_B{\bm\sigma}\cdot{\bm B} -e\Phi({\bm r})+\frac{\mu_B}{4mc^2}\Bigl[{\bm \nabla}\cdot{\bm E}\nonumber\\
&&+2{\bm E}\times({\bm p}+e{\bm A})\cdot{\bm \sigma}-i({\bm \nabla}\times{\bm E})\cdot{\bm \sigma}\Bigr].\label{H3D}\>\>\>
\end{eqnarray}
As noted long ago, for a spherically symmetric potential, ${\bm E}=-\hat{r}\frac{\partial V}{\partial r}$, the $({\bm E}\times{\bm p})\cdot{\bm\sigma}$ term reduces to spin-orbit coupling, such as that in the hydrogen atom \cite{BjorkenDrell}.  In addition, for a 3D conduction band, the $({\bm E}\times{\bm A})\cdot{\bm \sigma}$ term is  the quantum spin Hall contribution,  important for applied fields ${\bm E}||{\bm H}$ in thin films.
We remark that it is easy to show using the Klemm-Clem transformations of Eq. (\ref{HNR}) \cite{Klemmbook,KlemmClem}, that the Zeeman energy for a non-relativistic electron in an anisotropic 3D conduction band is precisely one-half of the Landau level splitting for all constant ${\bm B}$ directions.
\section{Low Dimensional Cases}
Recently, $H_{c2,||}(T)$ measurements were made  on monolayer or bilayer materials of transition-metal dichalcogenides such as MoS$_{2}$, WTe$_{2}$, NbSe$_{2}$ \cite{NbSe2,MoS2a,MoS2b,WTe2a,WTe2b,graphene2}. The results greatly exceeded the Pauli limit $\bm{B}_p$, even though the materials may  not be Ising superconductors \cite{WTe2a,WTe2b,graphene2}.

Here we present the limiting results from  Eq. (\ref{HNR}) for 2D  and 1D conduction bands. For 2D films or monolayers, we find
\begin{eqnarray}
{\cal H}_{2D}^{NR}&=&\frac{\Pi_1^2+\Pi_2^2}{2m}+\mu_B\sigma_{\perp}B_{\perp} -e\Phi({\bm r}_{||})+\frac{\mu_B}{4mc^2}\Bigl[{\bm \nabla}_{||}\cdot{\bm E}_{||}\nonumber\\
&&+2[{\bm E}\times({\bm p}+e{\bm A})]_{\perp}\sigma_{\perp}-i[{\bm \nabla}\times{\bm E}]_{\perp}\sigma_{\perp}\Bigr].\label{H2D}\>\>\>
\end{eqnarray}
We note that for a 2D thin film, the only component of the Zeeman energy is for the magnetic field normal to the film. For applied fields ${\bm E}||{\bm H}$, ${\bm A}$ leads to the quantum spin Hall effect. However, in 1D conductors, the non-relativistic Hamiltonian reduces to
\begin{eqnarray}
{\cal H}_{1D}^{NR}&=&\frac{\Pi_{||}^2}{2m}-e\Phi(r_{||})+\frac{\mu_B}{4mc^2}\frac{\partial E_{||}}{\partial x_{||}},\label{H1D}
\end{eqnarray}
where  $({\bm \Pi}_{1}, {\bm \Pi}_{2}, {\bm r}_{||})$ and $(\Pi_{||}, x_{||})$ respectively contain  the components of ${\bm\Pi}$, defined following Eq. (\ref{HNR}), and ${\bm r}$ in 2, and 1 dimensions. In 2D, the vector product can only occur for the two vectors in the conducting plane,  so the only component of the Zeeman interaction is  normal to the 2D conduction plane. In 1D, there is no vector product so there is no Zeeman interaction, as correctly omitted in phenomenological ${\bm H}_{c2}$ calculations \cite{TurkevichKlemm} .

These results have drastic consequences for the upper critical field ${\bm H}_{c2}$ parallel to 2D thin film superconductors, especially in the clean limit, for which strong spin-orbit scattering is unlikely to be able to  explain the large  violations of the Pauli limit \cite{KLB}.  The results are also profoundly important in the interpretation of Knight shift measurements for applied magnetic fields parallel to the layers of a layered superconductor or in any direction of a quasi-one-dimensional superconductor.  Without a Zeeman interaction, there is no solid reason to believe that the resonance frequency in such metals would be different from that in the insulating state.

After this work was completed, we found that  Eq. (\ref{HNR}) was derived previously using a different but equivalent procedure. That work correctly noted that there would be no Zeeman interaction in 1D, and  the Zeeman interaction in 2D would only be normal to the conducting plane \cite{Safonov}.   But that paper did not discuss any aspects of the covariance of the model, and   also did not include the ${\bm E}\times{\bm A}\cdot{\sigma}$ term in Eq. (\ref{HNR}).
\section{Conclusions}
The Dirac equation  is extended to treat a relativistic electron in an orthorhombically anisotropic conduction band. The norm for this model with metric $\tilde{g}$ is invariant under the most general proper Lorentz transformation $\tilde{A}$, the matrix representation of which exhibits  O(3,1) group symmetry, and this anisotropic Dirac equation is demonstrated to be covariant, precisely as for the isotropic Dirac equation. This model applies to large classes of anisotropic semiconductors, metals, and superconductors.  The orbital and spin components of the 3D non-relativistic Hamiltonian are intimately connected.

This model has  profound consequences for  the temperature dependence of  Knight shift measurements  and Pauli limiting effects upon ${\bm H}_{c2}$  for ${\bm H}$ parallel to the low mass direction(s) of  clean, highly anisotropic superconductors.  We encourage measurements at higher fields and lower $T$ values to confirm our prediction that ${\bm H}_{c2,||}(0)$ could greatly exceed the standard Pauli limit in clean monolayer and bilayer superconductors, such as gated and pure transition metal dichalcogenides and twisted bilayer graphene \cite{NbSe2,MoS2a,MoS2b,WTe2a,WTe2b,graphene2}.  Phenomenologically, the temperature dependence of $H_{c2,||}(T)$ in monolayer superconductors should behave as in the Tinkham thin film model \cite{Klemmbook}, but a microscopic theory that does not involve spin-orbit scattering or Ising pairing of $H_{c2,||}(T)$ in a clean two-dimensional superconductor is sorely needed \cite{KLB,NbSe2,MoS2b}.
\section{Acknowledgments}
  This work was supported by the National Natural Science Foundation of China through Grant no. 11874083.  AZ was also supported by the China Scholarship Council.

\end{document}